\documentclass[aps,twocolumn,prl,superscriptaddress]{revtex4}

\usepackage{graphicx}
\usepackage{amsmath}
\usepackage{amssymb}
\usepackage{units}
\usepackage{times}

\renewcommand{\vec}[1]{\boldsymbol{#1}}

\begin{document}

%
%

\title{Illuminating the dark corridor in graphene: polarization dependence of angle-resolved photoemission spectroscopy on graphene}

\author{Isabella Gierz}
\email[Corresponding author; electronic address:\
]{i.gierz@fkf.mpg.de} \affiliation{Max-Planck-Institut f\"ur
Festk\"orperforschung, 70569 Stuttgart, Germany}
\author{J\"urgen Henk}
\affiliation{Max-Planck-Institut f\"ur Mikrostrukturphysik, 06120 Halle (Saale), Germany}
\author{Hartmut H\"ochst}
\affiliation{Synchrotron Radiation Center, University of Wisconsin-Madison, Stoughton, WI 53589, USA}
\author{Christian R. Ast}
\affiliation{Max-Planck-Institut f\"ur Festk\"orperforschung,
70569 Stuttgart, Germany}
\author{Klaus Kern}
\affiliation{Max-Planck-Institut f\"ur Festk\"orperforschung,
70569 Stuttgart, Germany} \affiliation{IPMC, Ecole Polytechnique
F{\'e}d{\'e}rale de Lausanne, 1015 Lausanne, Switzerland}

\date{\today}

\begin{abstract}
We have used s- and p-polarized synchrotron radiation to image the electronic structure of epitaxial graphene near the $\overline{\text{K}}$-point by angular resolved photoemission spectroscopy (ARPES). Part of the experimental Fermi surface is suppressed due to the interference of photoelectrons emitted from the two equivalent carbon atoms per unit cell of graphene's honeycomb lattice. Here we show that by rotating the polarization vector, we are able to illuminate this `dark corridor' indicating that the present theoretical understanding is oversimplified. Our measurements are supported by first-principles photoemission calculations, which reveal that the observed effect persists in the low photon energy regime.
\end{abstract}

\maketitle

%
%


\begin{figure}
  \includegraphics[width = 1\columnwidth]{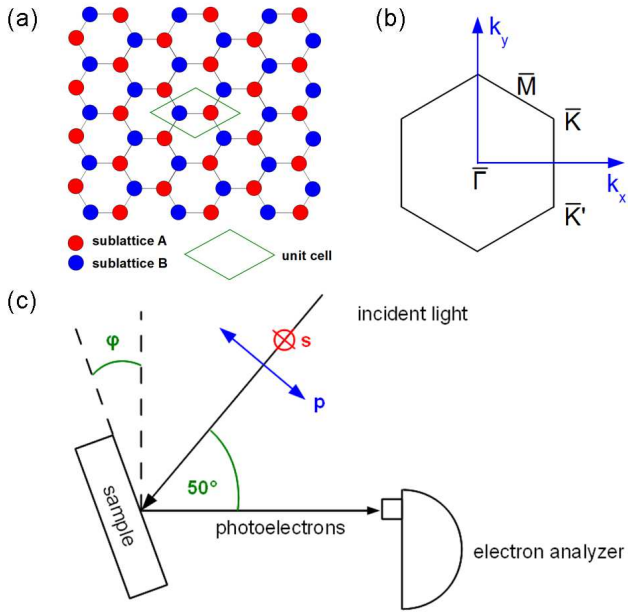}
  \caption{(Color online) honeycomb lattice with two equivalent carbon atoms per unit cell (a) together with the two-dimensional Brillouin zone (b). Panel (c) shows a sketch of the experimental setup. $k_y$ corresponds to a rotation of the sample around $\phi$. $k_x$ is the direction perpendicular to the paper plane, it corresponds to the dispersion direction in the 2D detector. For s(p)-polarized light the electric field vector lies perpendicular to the plane of incidence (in the plane of incidence) spanned by the sample normal and the direction of incidence of the light.}
  \label{figure1}
\end{figure}

Graphene, a single layer of sp$^2$-bonded carbon atoms, is one of the paradigm two-dimensional (2D) electron systems existing today. It is renowned for its high crystalline quality, its extremely high carrier mobility \cite{Novoselov,Berger2,Orlita} as well as its peculiar charge carriers that behave like massless Dirac particles \cite{Novoselov2,Zhang,Berger2,Geim,Bostwick1,Sprinkle} due to its honeycomb lattice consisting of two equivalent triangular sublattices A and B (see Fig. \ref{figure1}a). This leads to the description of graphene's charge carriers in terms of spinor wavefunctions in analogy to the Dirac equation for massless particles, where the `spin' index indicates the sublattice rather than the real electron spin, hence the term `pseudospin' \cite{Geim}. This pseudospin is responsible for graphene's many intriguing electronic properties. First of all, the difference in pseudospin of the two cosine-shaped bands originating from the two sublattices allows them to cross at the $\overline{\mbox{K}}$-point of the 2D Brillouin zone (see Fig. \ref{figure1}b) where they form the conical band structure \cite{Wallace,Slonczewski}. Second, due to the pseudospin the charge carriers accumulate a Berry phase of $\pi$ on closed loop paths resulting in the absence of backscattering. This has been observed in both magnetotransport \cite{Ando,Bliokh,Morozov,Wu} as well as scanning tunneling spectroscopy experiments \cite{Brihuega}. Furthermore, the pseudospin is responsible for the peculiar half-integer quantum Hall effect observed in graphene \cite{Novoselov2,Zhang,Wu2}. In addition, the conservation of the pseudospin upon passing a potential barrier is expected to result in perfect transparency of the barrier for graphene's charge carriers (Klein tunneling) \cite{Katsnelson}. The pseudospin concept has spawned ideas for different `pseudospintronic' device proposals, like e.g. the pseudospin valve \cite{pseudospin_valve}.

The effect of the pseudospin is also observed in angle-resolved photoemission spectroscopy (ARPES) experiments. Here, it is rather unwanted because it suppresses the photoemission intensity on part of the Fermi surface (`dark corridor' \cite{Shirley,Daimon,Kuemmeth}). The effect was verified many times in ARPES experiments using p-polarized light \cite{Bostwick1,Seyller,Mucha,Bostwick,Gierz1} and the presence of this dark corridor was never questioned. Unfortunately, the dark corridor effectively prevents the experimental verification of the spin rotation upon quasiparticle to photoelectron conversion in graphene, because of the lack of photoemission intensity in the region of interest \cite{Kuemmeth}.

Here we show that by using s-polarized light it is possible to illuminate this dark corridor and thereby access the complete Fermi surface of graphene in an ARPES experiment. While the dark corridor has been addressed theoretically before \cite{Shirley,Daimon} the polarization dependence of the intensity modulation on the Fermi surface cannot be accounted for by the single free-electron final state used in this model. We show that this problem is overcome in our first principles photoemission calculations where we use time-reversed spin-polarized low energy electron diffraction (SPLEED) states as final states.


A sketch of the experimental setup is displayed in Fig. \ref{figure1}c.
The measurements were done at the Synchrotron Radiation Center (SRC) in Stoughton, WI at the variable polarization VLS-PGM beamline. This beamline is equipped with an elliptically polarized Apple II undulator that delivers p- and s-polarization of photons in an energy range from 15\,eV to 250\,eV. For s(p)-polarized light the electric field vector lies perpendicular to the plane of incidence (in the plane of incidence) spanned by the sample normal and the direction of incidence of the light. For the ARPES experiments a Scienta analyzer with an energy resolution of better than 10\,meV was used. In order to measure the photoemission current as a function of $k_y$ the sample was rotated by an angle $\phi$ (see Fig. \ref{figure1}c) which was varied around $\phi_0=36.7^{\circ}$ for $h\nu=35$\,eV and around $\phi_0=28.7^{\circ}$ for $h\nu=52$\,eV. $k_x$ (direction perpendicular to the paper plane in Fig. \ref{figure1}c) corresponds to the dispersion direction in the 2D detector. During measurements the sample was kept at a temperature of 50K. We have grown graphene by thermal decomposition of SiC(0001) in ultra high vacuum \cite{Berger,Emtsev}. Details of the sample preparation are reported in References \cite{Gierz1} and \cite{Gierz0}.

First-principles electronic-structure calculations have been performed for a free-standing graphene layer, within the framework of relativistic multiple-scattering theory (layer Korringa-Kohn-Rostoker method \cite{Henk02,Zabloudil05}) using the Perdew-Wang exchange-correlation potential \cite{Perdew92}. The self-consistent potentials serve as input for the photoemission calculations, which rely on the relativistic one-step model \cite{Braun96,Henk02}. Thus, all essential ingredients of the excitation process are captured, in particular transition matrix elements and boundary conditions. Many-body effects are incorporated via the complex self-energy $\Sigma$. The imaginary part of $\Sigma$ is taken as 1.5 eV for the final state (time-reversed SPLEED state) and as 0.01 eV for the initial state (graphene orbitals); its real part is assumed zero. Including a non-zero real part of the self-energy would shift the final states to higher energies. Furthermore, the final state in experiment is scattered by the SiC substrate, so that deviations between the theoretical and the experimental final state are possible. These deviations may include slight changes in the final state composition as well as the band dispersion. Nevertheless, trends in experiment are fully accounted for, in particular the photon energy dependence of the intensities. For a direct comparison between experiment and theory the theoretical photon energies $h\nu_{\text{th}}$ have been shifted by 8.6\,eV towards higher photon energies.


\begin{figure}
  \includegraphics[width = 1\columnwidth]{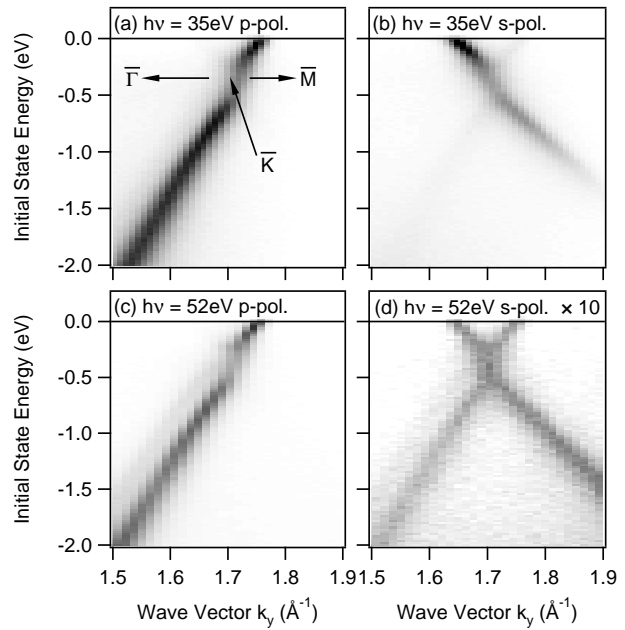}
  \caption{Band structure measured along $\overline{\Gamma\text{K}}$ for an epitaxial graphene monolayer on SiC(0001) for two different photon energies (a,b: 35\,eV; c,d: 52\,eV) for both p- (a,c) and s-polarized (b,d) light. The grey scale is linear with black (white) corresponding to high (low) photoemission intensities.}
  \label{figure2}
\end{figure}

Figure \ref{figure2} shows the measured band structure for an epitaxial graphene monolayer on SiC(0001) along the $\overline{\Gamma\text{KM}}$-direction. As epitaxial graphene on SiC(0001) is slightly n-doped due to charge transfer from the substrate, the crossing point of the two linearly dispersing $\pi$-bands is located at about $-420$\,meV below the Fermi level \cite{Bostwick1,Seyller,Mucha,Bostwick,Gierz1}. The data in Fig.\ \ref{figure2} was recorded at a photon energy of $h\nu=35$\,eV and $h\nu=52$\,eV with p- and s-polarized light. The grey scale is linear with black (white) corresponding to high (low) photoemission intensities. For p-polarized photons (Fig.\ \ref{figure2}a, c) the intensity for one of the two branches is completely suppressed due to interference effects in the photoemission process \cite{Shirley,Daimon}, only the branch dispersing upwards (towards the Fermi level) along $\overline{\Gamma\text{KM}}$ is visible in agreement with previous photoemission results \cite{Bostwick1,Seyller,Mucha,Bostwick,Gierz1}. For $h\nu=35$\,eV and s-polarized light (Fig.\ \ref{figure2}b) the photoemission intensity shifts to the second branch dispersing downwards (away from from the Fermi level) along $\overline{\Gamma\text{KM}}$ that was invisible when using p-polarized light. When using s-polarized light at $h\nu=52$\,eV (Fig.\ \ref{figure2}d) both $\pi$-bands are visible. In this case the overall intensity is reduced by about one order of magnitude as compared to the other measurements.

\begin{figure}
  \includegraphics[width = 1\columnwidth]{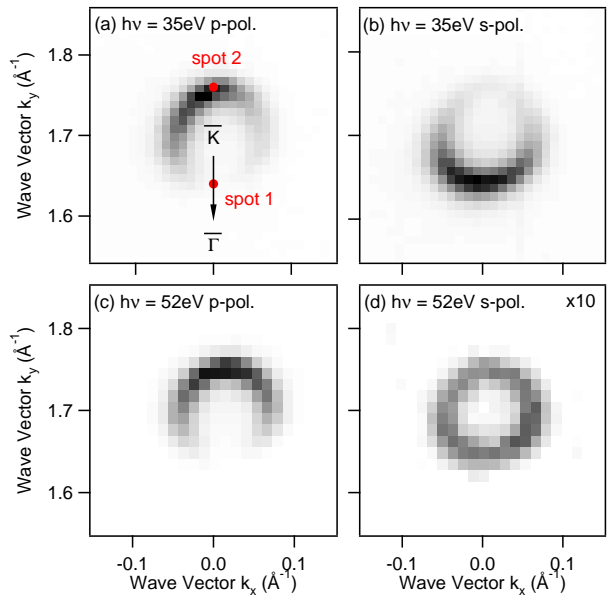}
  \caption{Fermi surface of epitaxial graphene on SiC(0001) measured with p-polarized light (a,c) and s-polarized light (b,d) for two different photon energies (a,b: 35\,eV; c,d: 52\,eV). The grey scale is linear with black (white) corresponding to high (low) photoemission intensities.}
  \label{figure3}
\end{figure}

Figure \ref{figure3} shows the corresponding Fermi surfaces around $\overline{\text{K}}$ for $h\nu=35$\,eV and $h\nu=52$\,eV with both p-polarized and s-polarized light. For p-polarized radiation (Fig.\ref{figure3}a,c) there is no photo\-emission intensity at spot 1. This situation changes drastically when using s-polarized photons with $h\nu=35$\,eV in Fig. \ref{figure3}b. In this case, there is no photoemission intensity at the opposite side of the Fermi surface at spot 2. Changing the photon energy to $h\nu=52$\,eV leads to a homogeneous illumination of the complete Fermi surface with s-polarized light (Fig. \ref{figure3}d). As in Fig. \ref{figure2}d, the photocurrent is one order of magnitude lower than for p-polarized radiation. As can be seen, the dark corridor at spot 1 as introduced by Refs. \cite{Shirley,Daimon,Kuemmeth} can be illuminated using s-polarized light.

The origin of the dark corridor has been explained by calculating the photoemission matrix element in dipole approximation using atomic orbitals for the initial state and a single plane wave for the final state \cite{Shirley}. It has been shown that the photoemission intensity around $\overline{\text{K}}$ can be separated into a polarization factor and an interference term related to the crystal structure. The interference term is responsible for the suppression of the photocurrent at spot 1 at the Fermi energy. The polarization factor ($\vec{k}\hat{\lambda}$) implies that the photoemission intensity vanishes completely for $\vec{k}\perp\hat{\lambda}$, i.\ e.\ for s-polarized radiation.

\begin{figure}
  \includegraphics[width = 1\columnwidth]{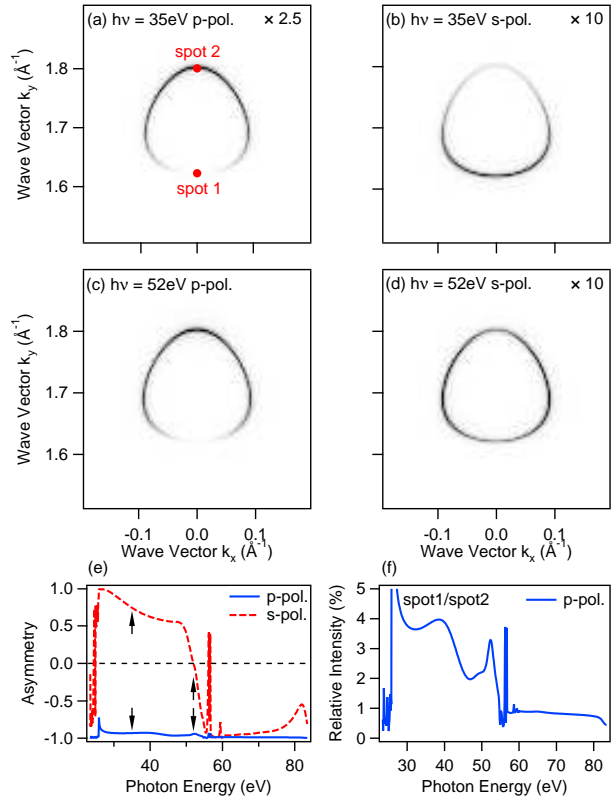}
  \caption{(Color online) Photoemission calculations of the Fermi surfaces for p-polarized (a,c) and s-polarized (b,d) radiation for $h\nu=35$\,eV (a,b) and $h\nu=52$\,eV (c,d). Panel (e) shows the intensity asymmetry of spot 1 and spot 2 as a function of photon energy (blue/continuous: p-pol.\ light; red/dashed: s-pol.\ light).  Panel (f) shows the intensity ratio of spot 1 compared to spot 2 for p-polarized light as a function of photon energy. The theoretical photon energies have been shifted by 8.6\,eV to allow for a direct comparison with experiment.}
  \label{figure4}
\end{figure}

However, our results show that this simple picture does not hold. For better agreement with the experimental findings we have used time-reversed SPLEED states as final states. Figure\ \ref{figure4} shows the calculated Fermi surface for p-polarized and s-polarized light with $h\nu=35$\,eV and $h\nu=52$\,eV. The calculation is in good agreement with the experimental results in Fig.\ \ref{figure3}. The dark corridor lies at spot 1 (spot 2) for p-polarized (s-polarized $h\nu=35$\,eV) light. For s-polarized light at $h\nu=52$\,eV the Fermi surface is completely illuminated. To complete the picture, Fig. \ref{figure4}e shows the intensity asymmetry between spot 1 and spot 2 defined as $A=(I_{spot 1}-I_{spot 2})/(I_{spot 1}+I_{spot 2})$ as a function of photon energy. For $A=\pm 1$, the dark corridor lies at spot 1 or spot 2, respectively. For $A=0$, spot 1 and spot 2 are equally illuminated, which is the case for $h\nu=52\,$eV and s-polarized light. The effect that spot 1 can be illuminated using s-polarized light persists for photon energies between $h\nu=24$\,eV and $h\nu=52$\,eV. The disappearance of the effect for $h\nu>52$\,eV is attributed to a change in the final states. Decomposing the time-reversed SPLEED final states into angular-momentum partial waves, we find that for $h\nu<52$\,eV s-like partial waves dominate the photoemission process while for $h\nu>52$\,eV the contributions from d-like partial waves dominate.

In order to compare our calculations with the results from Ref.\ \cite{Shirley}, we project the time-reversed SPLEED final states onto free-electron final states. This decomposition shows that the photoemission process is dominated by up to twelve different plane waves in contrast to the single plane wave used in \cite{Shirley}. The weight of the different plane waves depends on the photon energy. As for the partial wave decomposition there is a transition between different plane wave contributions around $h\nu=52$\,eV. Our plane wave decomposition reveals that the plane wave $e^{i\vec{k}\vec{x}}$ used in Ref. \cite{Shirley} contributes at all photon energies. This explains the success of the model for p-polarized light. However, in order to explain the experimental results for s-polarized light within a tight-binding calculation, it is necessary to employ more than just one plane wave final state. Detailed calculations are given as EPAPS.

Fig.\ \ref{figure4}f shows the relative intensity of spot 1 compared to spot 2 as a function of photon energy. The photoemission intensity at spot 1 does not go to zero but remains at a few percent for p-polarized light, even though perfect AB sublattice symmetry is assumed. This is in contrast to the tight-binding calculation in Ref.\ \cite{Shirley}, where perfect AB sublattice symmetry leads to zero intensity in the dark corridor. This discrepancy can be understood by including the spin-orbit interaction (SOI) in the tight-binding model (see EPAPS). As a result, the wave function coefficients $c_A$ and $c_B$ of the $p_z$-orbitals centered at the A and B sublattice, respectively, are not equal in magnitude anymore, which leads to a nonzero photocurrent inside the dark corridor. As a consequence, the degree of AB sublattice symmetry breaking cannot be deduced from the intensity inside the dark corridor as was suggested in Refs.\ \cite{Seyller,Bostwick}, unless the influence  of the SOI is precisely known. Nevertheless, as the SOI in graphene is small, the same is to be expected for the corresponding photoemission intensity.

Furthermore, Ref. \cite{Kuemmeth} predicts a giant spin rotation during quasiparticle to photoelectron conversion in graphene due to spin-pseudospin interference in the photoemission process. Inside the dark corridor (at spot 1) the spin orientation of the photoelectron differs from the spin of the quasiparticle in the initial state by $180^{\circ}$. However, up to now this effect was believed not to be accessible in a spin-resolved ARPES measurement because of the lack of photoemission intensity inside the region of interest. Using s-polarized radiation in a spin-resolved ARPES experiment should allow for the experimental verification of the predicted spin rotation.


In conclusion, we could show that it is possible to illuminate the dark corridor on the measured Fermi surface of graphene using s-polarized synchrotron radiation. This effect is not included in the theoretical model from Ref. \cite{Shirley} that is based on a single free electron final states. Our first principles photoemission calculations use time-reversed SPLEED states as final states and result in good agreement with the measured Fermi surfaces. In addition, the calculations reveal that the observed effect persists in the low photon energy regime up to about $h\nu=52$\,eV. Furthermore, our findings open up a new pathway to access the giant spin rotation predicted in Ref. \cite{Kuemmeth} experimentally in a spin-resolved ARPES measurement.

The authors thank U.\ Starke and C.\ Riedl
from the Max Planck Institute for Solid State Research in Stuttgart as
well as C.\ L.\ Frewin, C.\ Locke and S.\ E.\ Saddow of the University
of South Florida for hydrogen etching of the SiC substrates. C.\ R.\ A.\ acknowledges funding by the
Emmy-Noether-Program of the Deutsche Forschungsgemeinschaft (DFG).
This work is based in part upon research conducted at the
Synchrotron Radiation Center of the University of
Wisconsin-Madison which is funded by the National Science
Foundation under Award No DMR-0537588.

\end{document}